\documentclass[11pt]{article}
\usepackage{graphicx}
\usepackage{epsf}  

\newcommand{\psfile}[3][]{ 
  \begin{center}
    \setlength{\epsfxsize}{#3\linewidth}\leavevmode
    \def\noOpt{}\def\testit{#1}\ifx\testit\noOpt%
      \epsfbox{#2}%
    \else%
      \epsfbox[#1]{#2}%
    \fi
  \end{center} 
}

\newcommand{\BABARPubYear}    {03}

\newcommand{\BABARConfNumber} {017}
\newcommand{\SLACPubNumber} {10099}

\input pubboard/babarsym

\newcommand{\pvec}{{\bf p}}

\newcommand{\acp}{\ensuremath{\calA_{ch}}}
\newcommand{\calB}{\ensuremath{{\cal B}}}

\newcommand{\DE}{\ensuremath{\Delta E}}

\newcommand{\xf}{\ensuremath{{\cal F}}}
\newcommand{\hel}{\ensuremath{{\cal H}}}
\newcommand{\costhr}{\ensuremath{\cos\theta_{\rm T}}}

\newcommand\etal{{\it et al.}}
\newcommand{\half}{\ensuremath{{1\over2}}}

\newcommand{\bfig}{\begin{figure}[htbpc!]}
\newcommand{\efig}{\end{figure}}
\newcommand\bef{\begin{figure}}
\newcommand\edf{\end{figure}}
\newcommand\dbline{\noalign{\vskip 0.10truecm\hrule}\noalign{\vskip 2pt}\noalign{\hrule\vskip 0.10truecm}}
\newcommand\sgline{\noalign{\vskip 0.10truecm\hrule\vskip 0.10truecm}}
\newcommand\beq{\begin{equation}}
\newcommand\eeq{\end{equation}}
\newcommand\bear{\begin{array}}
\newcommand\enar{\end{array}}
\newcommand\beqa{\begin{eqnarray}}
\newcommand\eeqa{\end{eqnarray}}
\newcommand\ben{\begin{enumerate}}
\newcommand\een{\end{enumerate}}

\newcommand{\UfourS}{\ensuremath{\Upsilon(4S)}}

\newcommand{\etagg}{\ensuremath{\eta_{\gaga}}}
\newcommand{\etappp}{\ensuremath{\eta_{3\pi}}}
\newcommand{\etatogg}{\ensuremath{\eta\ra\gaga}}
\newcommand{\etatoppp}{\ensuremath{\eta\ra\pi^+\pi^-\pi^0}}

\newcommand{\etapepp}{\ensuremath{\etapr_{\eta\pi\pi}}}
\newcommand{\etaptoepp}{\ensuremath{\etapr\ra\eta\pip\pim}}
\newcommand{\etaprg}{\ensuremath{\etapr_{\rho\gamma}}}
\newcommand{\etaptorg}{\ensuremath{\etapr\ra\rho^0\gamma}}

\newcommand{\etaprp}{\ensuremath{\eta^{(\prime)}}}		

\newcommand{\Kst}{\ensuremath{K^*}}

   \newcommand{\KstpKppiz}{\ensuremath{\Kstarp_{K^+\pi^0}}}
   
   \newcommand{\KstpKspip}{\ensuremath{\Kstarp_{\KS\pi^+}}}
   
   \newcommand{\KstzKppim}{\ensuremath{\Kstarz_{K^+\pi^-}}}

   \newcommand{\rhop}{\ensuremath{\rho^+}}
   
\newcommand{\kzs}{\ensuremath{K^0_S}}

\newcommand{\Dcontrol}{\ensuremath{\Bm\ra\pim\Dz~{\rm and}~\Bm\ra\rho^-\Dz~{\rm with}~\Dz\ra K^{-}\pip\piz}}

\newcommand{\fetaK}{\ensuremath{\eta K}}
\newcommand{\etaK}{\ensuremath{\B\ra\fetaK}}

\newcommand{\fetaKst}{\ensuremath{\eta K^{*}}}
\newcommand{\etaKst}{\ensuremath{\B\ra\fetaKst}}

\newcommand{\fetaKstp}{\ensuremath{\eta K^{*+}}}
\newcommand{\etaKstp}{\ensuremath{\Bp\ra\fetaKstp}}
\newcommand{\BetaKstp}{\ensuremath{\calB(\etaKstp)}}
\newcommand{\retaKstp}{\ensuremath{xx^{+xx}_{-xx}\pm xx}}
\newcommand{\RetaKstp}{\ensuremath{(\retaKstp\,)\times 10^{-6}}}
\newcommand{\AetaKstp}{\ensuremath{xx\pm xx\pm xx}}
\newcommand{\setaKstp}{\ensuremath{xx}}
  \newcommand{\fetaggKstpKspip}{\ensuremath{\eta_{\gamma\gamma}K^{*+}_{\KS \pi^+}}}
  
  \newcommand{\fetaggKstpKppiz}{\ensuremath{\eta_{\gamma\gamma} K^{*+}_{K^+\pi^0}}}
  
  \newcommand{\fetapppKstpKspip}{\ensuremath{\eta_{3\pi}K^{*+}_{\KS \pi^+}}}
  
  \newcommand{\fetapppKstpKppiz}{\ensuremath{\eta_{3\pi}K^{*+}_{K^+\pi^0}}}

\newcommand{\fetaKstz}{\ensuremath{\eta K^{*0}}}
\newcommand{\etaKstz}{\ensuremath{\Bz\ra\fetaKstz}}
\newcommand{\BetaKstz}{\ensuremath{\calB(\etaKstz)}}
\newcommand{\retaKstz}{\ensuremath{xx^{+xx}_{-xx}\pm xx}}
\newcommand{\RetaKstz}{\ensuremath{(\retaKstz)\times 10^{-6}}}
\newcommand{\AetaKstz}{\ensuremath{xx\pm xx \pm xx}}
\newcommand{\setaKstz}{\ensuremath{xx}}
  \newcommand{\fetaggKstz}{\ensuremath{\eta_{\gamma\gamma}K^{*0}}}

  \newcommand{\fetapppKstz}{\ensuremath{\eta_{3\pi} K^{*0}}}

\newcommand{\fetarho}{\ensuremath{\eta\rho}}

\newcommand{\fetarhop}{\ensuremath{\eta\rho^+}}
\newcommand{\etarhop}{\ensuremath{\Bp\ra\fetarhop}}
\newcommand{\Betarhop}{\ensuremath{\calB(\etarhop)}}
\newcommand{\retarhop}{\ensuremath{xx^{+xx}_{-xx}\pm xx}}
\newcommand{\Retarhop}{\ensuremath{(\retarhop)\times 10^{-6}}}
\newcommand{\Aetarhop}{\ensuremath{xx\pm xx \pm xx}}
\newcommand{\setarhop}{\ensuremath{xx}}
  \newcommand{\fetaggrhop}{\ensuremath{\eta_{\gamma\gamma} \rho^+}}
  
  \newcommand{\fetappprhop}{\ensuremath{\eta_{3\pi} \rho^+}}

\newcommand{\fetappi}{\ensuremath{\etapr\pi}}
\newcommand{\etappi}{\ensuremath{\B\ra\fetappi}}
\newcommand{\fetapK}{\ensuremath{\etapr K}}
\newcommand{\etapK}{\ensuremath{\B\ra\fetapK}}

\newcommand{\fetappip}{\ensuremath{\etapr\pip}}
\newcommand{\etappip}{\ensuremath{\Bp\ra\fetappip}}
\newcommand{\Betappip}{\ensuremath{\calB(\Bp\ra\etapr \pip)}}
\newcommand{\retappip}{\ensuremath{xx^{+xx}_{-xx} \pm xx}}
\newcommand{\Retappip}{\ensuremath{(\retappip)\times 10^{-6}}}
\newcommand{\uletappip}{\ensuremath{xx}}
\newcommand{\ULetappip}{\ensuremath{\uletappip\times 10^{-6}}}

\newcommand{\setappip}{\ensuremath{xx}}
   \newcommand{\fetapepppip}{\ensuremath{\etapr_{\eta\pi\pi} \pi^+}}
   
   \newcommand{\fetaprgpip}{\ensuremath{\etapr_{\rho\gamma} \pi^+}}

\newcommand{\fetapKp}{\ensuremath{\etapr K^+}}

\newcommand{\etapKst}{\ensuremath{\B\ra\etapr K^{*}}}

\newcommand{\fetapKstp}{\ensuremath{\etapr K^{*+}}}
\newcommand{\etapKstp}{\ensuremath{\Bp\ra\etapr K^{*+}}}
\newcommand{\BetapKstp}{\ensuremath{\calB(\etapKstp)}}
\newcommand{\retapKstp}{\ensuremath{xx^{+xx}_{-xx}\pm xx}}

\newcommand{\uletapKstp}{\ensuremath{xx}}
\newcommand{\ULetapKstp}{\ensuremath{\uletapKstp\times 10^{-6}}}
\newcommand{\setapKstp}{\ensuremath{xx}}

   \newcommand{\fetapeppKstpKspip}{\ensuremath{\etapr_{\eta\pi\pi}\Kstarp_{\KS\pip}}}
   
   \newcommand{\fetapeppKstpKppiz}{\ensuremath{\etapr_{\eta\pi\pi}\Kstarp_{\Kp\piz}}}
   
   \newcommand{\fetaprgKstp}{\ensuremath{\etapr_{\rho\gamma}\Kstarp}}
   
   \newcommand{\fetaprgKstpKspip}{\ensuremath{\etapr_{\rho\gamma}\Kstarp_{\KS\pip}}}
   
   \newcommand{\fetaprgKstpKppiz}{\ensuremath{\etapr_{\rho\gamma}\Kstarp_{\Kp\piz}}}

\newcommand{\fetapKstz}{\ensuremath{\etapr K^{*0}}}
\newcommand{\etapKstz}{\ensuremath{\Bz\ra\fetapKstz}}
\newcommand{\BetapKstz}{\ensuremath{\calB(\etapKstz)}}
\newcommand{\retapKstz}{\ensuremath{xx^{+xx}_{-xx}\pm xx}}

\newcommand{\uletapKstz}{\ensuremath{xx}}
\newcommand{\ULetapKstz}{\ensuremath{\uletapKstz\times 10^{-6}}}
\newcommand{\setapKstz}{\ensuremath{xx}}

   \newcommand{\fetapeppggKstz}{\ensuremath{\etapr_{\eta(\gamma\gamma)\pi\pi} \Kstarz}}
   
   \newcommand{\fetapfivepiKstz}{\ensuremath{\etapr_{\eta(3\pi)\pi\pi} \Kstarz}}
   
   \newcommand{\fetaprgKstz}{\ensuremath{\etapr_{\rho\gamma} K^{*0}}}

\newcommand{\fetaprhop}{\ensuremath{\etapr\rho^+}}
\newcommand{\etaprhop}{\ensuremath{\Bp\ra\fetaprhop}}
\newcommand{\Betaprhop}{\ensuremath{\calB(\etaprhop)}}
\newcommand{\retaprhop}{\ensuremath{xx^{+xx}_{-xx}\pm xx}}
\newcommand{\Retaprhop}{\ensuremath{(\retaprhop)\times 10^{-6}}}
\newcommand{\uletaprhop}{\ensuremath{xx}}
\newcommand{\ULetaprhop}{\ensuremath{\uletaprhop\times 10^{-6}}}

\newcommand{\setaprhop}{\ensuremath{xx}}
  \newcommand{\fetaprgrhop}{\ensuremath{\etapr_{\rho\gamma}\rhop}}
  \newcommand{\fetapepprhop}{\ensuremath{\etapr_{\eta\pi\pi}\rhop}}

\renewcommand{\retaKstp}{\ensuremath{25.7^{+3.8}_{-3.6}\pm1.8}}
\renewcommand{\AetaKstp}{\ensuremath{+0.15\pm0.14\pm0.02}}
\renewcommand{\setaKstp}{\ensuremath{12}}

\renewcommand{\retaKstz}{\ensuremath{19.0^{+2.2}_{-2.1}\pm 1.3}}
\renewcommand{\AetaKstz}{\ensuremath{+0.03\pm0.11\pm0.02}}
\renewcommand{\setaKstz}{\ensuremath{15}}

\renewcommand{\retarhop}{\ensuremath{10.5^{+3.1}_{-2.8}\pm1.3}}
\renewcommand{\Aetarhop}{\ensuremath{+0.06\pm0.29\pm0.02}}
\renewcommand{\setarhop}{\ensuremath{4.8}}

\renewcommand{\retappip}{\ensuremath{2.8^{+1.3}_{-1.0} \pm 0.3}}
\renewcommand{\uletappip}{\ensuremath{4.5}}

\renewcommand{\setappip}{\ensuremath{3.4}}

\renewcommand{\retapKstp}{\ensuremath{6.1^{+3.9}_{-3.2}\pm 1.2}}
\renewcommand{\uletapKstp}{\ensuremath{12}}
\renewcommand{\setapKstp}{\ensuremath{2.4}}

\renewcommand{\retapKstz}{\ensuremath{3.2^{+1.8}_{-1.6}\pm 0.9}}
\renewcommand{\uletapKstz}{\ensuremath{6.4}}
\renewcommand{\setapKstz}{\ensuremath{2.2}}

\renewcommand{\retaprhop}{\ensuremath{14.0^{+5.1}_{-4.6}\pm 1.9}}
\renewcommand{\uletaprhop}{\ensuremath{22}}

\renewcommand{\setaprhop}{\ensuremath{3.8}}

\long\def\inst#1{\par\nobreak\kern 4pt\nobreak
    {\it #1}\par\vskip 10pt plus 3pt minus 3pt}

\begin{document}
{\pagestyle{empty}


\begin{flushright}
\babar-CONF-\BABARPubYear/\BABARConfNumber \\
SLAC-PUB-\SLACPubNumber \\
August 2003 \\
\end{flushright}

\par\vskip 5cm

\begin{center}
\Large \bf\boldmath 
Measurement of Branching Fractions and Charge Asymmetries in \B\ Meson Decays to $\eta^{(\prime)}K^*$, $\eta^{(\prime)}\rho$, and $\eta^{\prime}\pi$ \\
\end{center}
\bigskip

\begin{center}
\large The \babar\ Collaboration\\
\mbox{ }\\
\today
\end{center}
\bigskip \bigskip

\begin{center}
\large \bf Abstract
\end{center}
We present preliminary measurements of branching fractions and charge
asymmetries for the $B$ meson decays 
$B\ra\eta^{(\prime)}K^*$, $B\ra\eta^{(\prime)}\rho$, and \etappip .
The data were recorded with the \babar\ detector 
at PEP-II and correspond to 
$89\times 10^6$ \BB\ pairs produced in \epem\ annihilation through the
\UfourS\ resonance.  We find the branching fractions 
$\BetaKstz = \RetaKstz$, $\BetaKstp = \RetaKstp$, $\Betarhop = \Retarhop$,
$\Betaprhop = \Retaprhop\ (< \ULetaprhop$ with $90\%$ confidence), and 
$\Betappip = \Retappip\ (< \ULetappip$).
We also set $90\%$ CL upper limits of 
$\BetapKstz < \ULetapKstz$ and
$\BetapKstp < \ULetapKstp$.
The time-integrated charge asymmetries are
$\acp(\fetaKstz)=\AetaKstz$, $\acp(\fetaKstp)=\AetaKstp$, and 
$\acp(\fetarhop)=\Aetarhop$.

\vfill
\begin{center}
Contributed to the 
XXI$^{\rm st}$ International Symposium on Lepton and Photon Interactions at High~Energies, 8/11 --- 8/16/2003, Fermilab, Illinois USA
\end{center}

\vspace{1.0cm}
\begin{center}
{\em Stanford Linear Accelerator Center, Stanford University, 
Stanford, CA 94309} \\ \vspace{0.1cm}\hrule\vspace{0.1cm}
Work supported in part by Department of Energy contract DE-AC03-76SF00515.
\end{center}

\newpage
} 

\begin{center}
\small

The \babar\ Collaboration,
\bigskip

%
B.~Aubert,
R.~Barate,
D.~Boutigny,
J.-M.~Gaillard,
A.~Hicheur,
Y.~Karyotakis,
J.~P.~Lees,
P.~Robbe,
V.~Tisserand,
A.~Zghiche
\inst{Laboratoire de Physique des Particules, F-74941 Annecy-le-Vieux, France }
% ...

\end{center}\newpage

\setcounter{footnote}{0}

%
%
%
\section{Introduction}\label{sec:intro}

We report the results of searches for $B$ decays to the charmless
final states\footnote{Except as noted explicitly, we use a particle name
to denote either member of a charge conjugate pair.}
$\etaprp\Kst$ , $\etaprp\rho$ , and \fetappip .  
For decays that are self tagging with respect to the $\b$ or $\bbar$ flavor
we also measure the direct \CP-violating time-integrated charge asymmetry, 
$\acp =(\Gamma^--\Gamma^+)/(\Gamma^-+\Gamma^+)$.
For charged \B\ decays $\Gamma^\pm\equiv\Gamma(B^\pm\ra
\eta^{(\prime)} h^\pm)$, and for \etaKstz\ with $\Kstarz\ra\Kp\pim$ the
sign on $\Gamma$ matches that of the tertiary kaon.  

\begin{figure}[!hb]
\scalebox{1.0}[1.0]{ \includegraphics{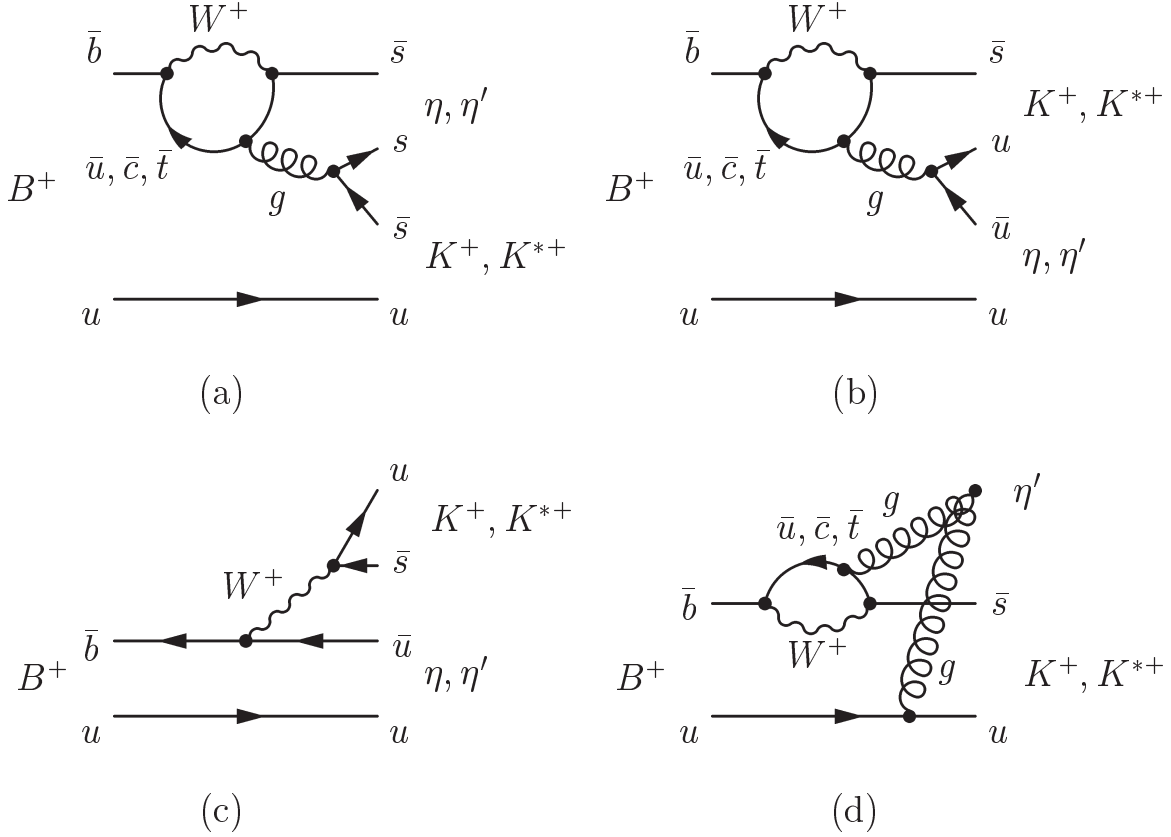}}
\caption{\label{fig:feyn_etakstar} Feynman diagrams for the decays
$B\ra (\eta,\etapr)(K,\Kstar)^+$.  The neutral decays are similar except that
the spectator quark becomes a $d$, and the tree diagram is internal.
}
\end{figure}

Interest in $B$ decays to $\eta$ or \etapr\ final states intensified in 1997
with the CLEO observation of the decay \etapK\ \cite{CLEOetapobs}.  It had
been pointed out by Lipkin six years earlier \cite{Lipkin} that
interference between two penguin diagrams (see Fig. \ref{fig:feyn_etakstar}a 
and \ref{fig:feyn_etakstar}b) and the known $\eta/\etapr$
mixing angle conspire to greatly enhance \etapK\ and suppress \etaK.  
Because the vector \Kstar\ has the opposite parity from the kaon,
the situation is reversed for the \etapKst\ and \etaKst\ decays.  The
general features of this picture have already been verified by previous 
measurements and limits.  However the details and possible contributions 
of flavor-singlet diagrams ({\it c.f.} Fig. \ref{fig:feyn_etakstar}d) 
can only be tested with the measurement 
of the branching fractions of all four $(\eta,\etapr)(K,\Kstar)$ decays;
the branching fraction of the \etapKst\ decay is expected to be particularly 
sensitive to a flavor-singlet component \cite{chiang,beneke}.  In any
case the tree diagrams (Fig. \ref{fig:feyn_etakstar}c) are CKM suppressed.
The results described in this paper complete the measurement of all
four decays with a \babar\ dataset of 89 million \BB\ decays
\cite{bbetapkPub,bbetakpiConf}. 

\begin{figure}[!hb]
\scalebox{1.0}[1.0]{ \includegraphics{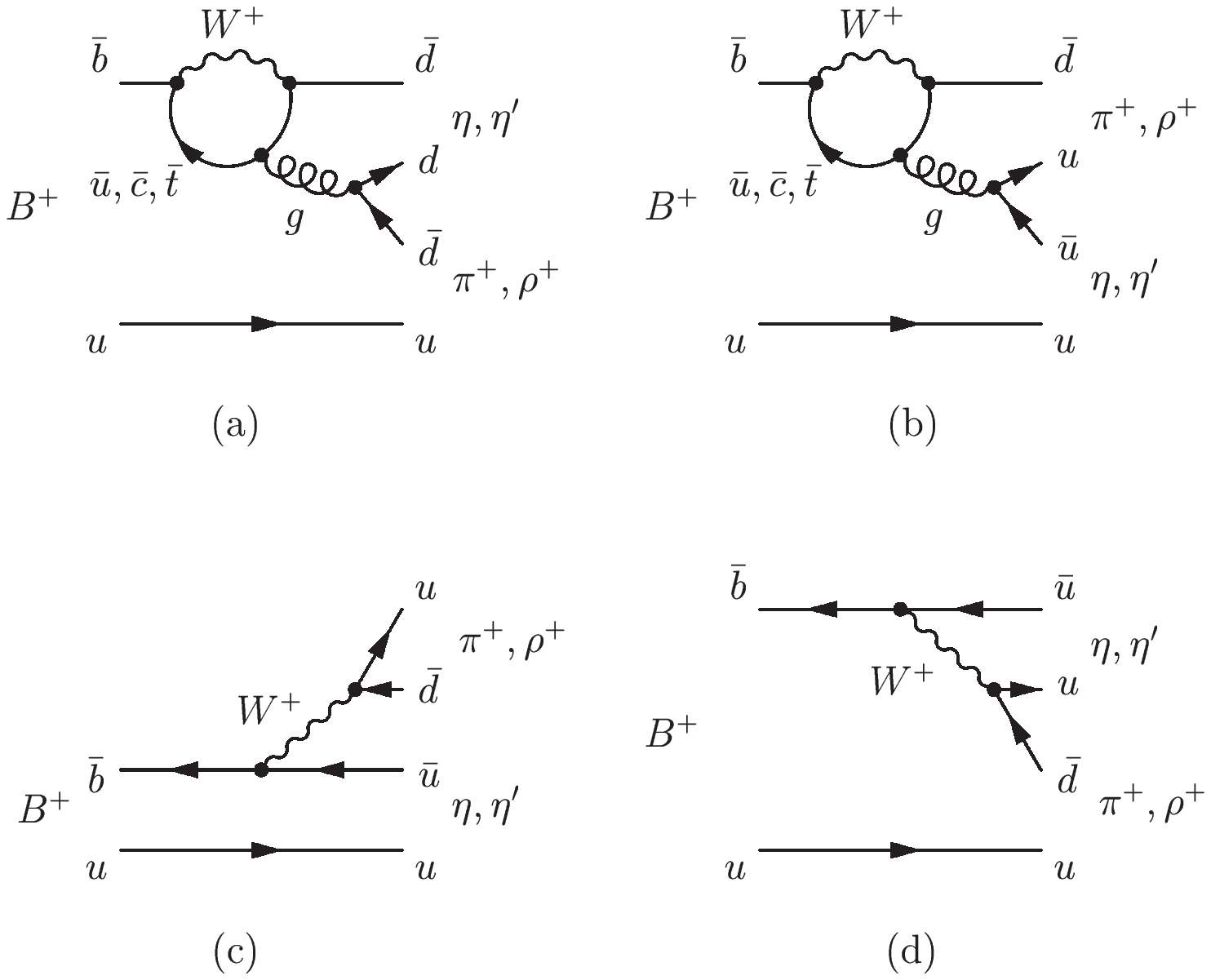}}
\caption{\label{fig:feyn_etapirho} Feynman diagrams for the decays
\etarhop, \etappip, and \etaprhop.  
}
\end{figure}

The situation for the decays $\etaprp\rho$ and \fetappip\ is different.
These decays are expected to be dominated by tree diagrams (see Fig.
\ref{fig:feyn_etapirho}c and \ref{fig:feyn_etapirho}d) since it is now the 
penguin diagrams (Fig. \ref{fig:feyn_etapirho}a and \ref{fig:feyn_etapirho}b) 
that are suppressed.  Since the internal tree diagram (Fig. 
\ref{fig:feyn_etapirho}d) is color suppressed, 
any interference effect is diluted for these decays.  Branching 
fractions for these decays are generally expected to be in the range 
(1--10)$\times10^{-6}$ \cite{kramer,grabbag,yang,CGW}.

The charge asymmetry \acp\ for most of these decays is expected to be small
\cite{kramer,AKLasym}.  However,
for the decay \etappi\ and \etapKst, the penguin and tree diagrams are expected 
to be of similar magnitude, which allows
potentially large charge asymmetries \cite{beneke,yang,CGW,dighe}.

The current knowledge of these decays comes from published measurements from
CLEO~\cite{CLEOetapr} and \babar\ \cite{BABARetapOmega}, and conference results 
from \babar\ \cite{BABAReta} and Belle \cite{Belleeta}.
Tables~\ref{tab:OldResultsa} and \ref{tab:OldResultsb} summarize these 
previous results.  


\begin{table}[htb]
\caption{Summary of branching fraction results for $B$ decays to $\eta$
    mesons from CLEO~\cite{CLEOetapr}, previous \babar
    ~\cite{BABAReta} measurements, Belle~\cite{Belleeta}, and the
    present analysis.  The results for all fits are given as well as a
    90\% CL upper limit if given in the cited reference.  The overall yields and efficiencies ($\epsilon$) are
    given as the sum of yields and efficiencies from the 
    daughter particle decay channels.  Rows marked (*) refer to
    measurements that are superseded by the results of the present
    paper.  }
\label{tab:OldResultsa}
\begin{center}
\vspace*{-0.3cm}
\hspace*{-0.5cm}
\begin{tabular}{l|ccccccc}
    \dbline
Expt.    &\# \BB\ ($10^{6}$)&Fit \calB$(10^{-6})$&UL \calB$(10^{-6})$&Signif. ($\sigma$)&Signal yield&$\epsilon$ (\%) \\
\sgline
\fetaKstz  & & & & & & \\
~~CLEO     &10	& $13.8^{+5.5}_{-4.6}\pm1.6$ 	& --- 	& 5.1 	& 15.6 	& 11.6 \\
~~\babar*   &23	& $19.1^{+6.3}_{-5.4}$		& ---	& 5.4 	&$21.7^{+7.1}_{-6.1}$ &4.9 \\
~~Belle    &23	& $21.2^{+5.4}_{-4.7}\pm2.0$ 	& --- 	& 5.1 	& 22.1 	& 4.7 	\\
This result&89  & \retaKstz			& ---	&\setaKstz& 157	& 9.2	\\
\sgline
\fetaKstp & & & & & & \\
~~CLEO     &10	& $26.4^{+9.6}_{-8.2}\pm3.3$ 	& --- 	& 4.8 	& 19.2 	  & 7.0 \\
~~\babar*  &23	& $22.1^{+11.1}_{-9.2}$ 	&$<34$	& 3.2 	&$14\pm7$ & 3.2 \\
~~Belle    &23	& ---				&$<50$  & 2.8 	& 13.8 	  & 2.5 \\
This result&89  & \retaKstp 			& ---	&\setaKstp& 113	  & 4.7	\\
\sgline
\fetarhop & & & & & & \\
~~CLEO     &10	& $4.8^{+5.2}_{-3.8}$ 		&$<15$ 	& 1.3 	& 5.0 	& 11.9 	\\
~~Belle    &23	& ---				&$<6.8$ & 0.0 	& 0.0 	& 3.9 	\\
This result&89  & \retarhop			& ---	&\setarhop& 65	& 7.0	\\
\dbline
\end{tabular}
\end{center}
\end{table}


\begin{table}[htb]
\caption{Summary of branching fraction results for $B$ decays to $\etapr$
    mesons from CLEO~\cite{CLEOetapr}, previous \babar ~\cite{BABAReta,BABARetapOmega} 
    measurements, Belle~\cite{Belleeta}, and the
    present analysis.  The results for all fits are given as well as a 90\%
    CL upper limit if  given in the cited reference.
    The overall yields and efficiencies ($\epsilon$) are given as the sum
    of yields and efficiencies from the daughter particle decay channels.
  Rows marked (*) refer to measurements that are superseded by
the results of the present paper.
}
\label{tab:OldResultsb}
\begin{center}
\vspace*{-0.3cm}
\hspace*{-0.5cm}
\begin{tabular}{l|ccccccc}
    \dbline
Expt.    &\# \BB\ ($10^{6}$)&Fit \calB$(10^{-6})$&UL \calB$(10^{-6})$&Signif. ($\sigma$)&Signal yield&$\epsilon$ (\%) \\
\sgline
\fetappip & & & & & & \\
~~CLEO     &10 & $0.1^{+0.5}_{-0.1}$ 		&$<11.1$ & 0.2 	& 4.4 	& 13.7 \\
~~\babar*  &23	& $5.4^{+3.5}_{-2.6} \pm 0.8$ & $<12$ 	 & 2.8 	& 5.7 & 9.1& \\
This result&89  & \retappip		&$<\uletappip$	 & 3.4	& 13	& 9.9 \\
\sgline
\fetapKstz & & & & & & \\
~~CLEO     &10  & $7.8^{+7.7}_{-5.7}$ 		&$<24$ 	 & 1.8 	& 2.4 	& 6.4 \\
~~\babar*  &54	& $4.0^{+3.5}_{-2.4}\pm1.0$ 	&$<13$ 	 & 1.8 	& $4.4^{+3.8}_{-2.6}$ & 1.9 \\
This result&89  & \retapKstz		&$<\uletapKstz$ &\setapKstz & 22 & 6.6 \\
\sgline
\fetapKstp  & & & & & & \\
~~CLEO     &10	& $11.1^{+12.7}_{-8.0}$ 	&$<35$ 	 & 2.5 	& 3.3 	 & 3.6 \\
This result&89  & \retapKstp		&$<\uletapKstp$  &\setapKstp &15 & 3.0 \\
\sgline
\fetaprhop   & & & & & & \\
~~CLEO     &10  & $11.2^{+11.9}_{-7.0}$ 	&$<33$ 	 & 2.4 	& 5.8 	 & 5.2 \\
This result&89  & \retaprhop		& $<\uletaprhop$ &\setaprhop& 69 & 4.1 \\
\dbline
\end{tabular}
\end{center}
\end{table}


%
%
\section{Detector and Data} \label{sec:detector}

The results presented in this paper are based on data collected
with the \babar\ detector~\cite{BABARNIM}
at the PEP-II asymmetric $e^+e^-$ collider~\cite{pep}
located at the Stanford Linear Accelerator Center.  An integrated
luminosity of 82~\invfb, corresponding to 
89 million \BB\ pairs, 
was recorded at the $\Upsilon (4S)$ resonance
(``on-resonance'', center-of-mass energy $\sqrt{s}=10.58\ \gev$).
An additional 9.6~\invfb\ were taken about 40~MeV below
this energy (``off-resonance'') for the study of continuum backgrounds in
which a light or charm quark pair is produced instead of an \UfourS.

The asymmetric beam configuration in the laboratory frame
provides a boost of $\beta\gamma = 0.56$ to the $\Upsilon(4S)$.
Charged particles (tracks) are detected and their momenta measured by the
combination of a silicon vertex tracker (SVT), consisting of five layers
of double-sided detectors, and a 40-layer central drift chamber,
both operating in the 1.5-T magnetic field of a solenoid. Photons and
electrons (neutral clusters) are detected by a CsI(Tl) electromagnetic
calorimeter (EMC). 

Charged-particle identification (PID) is provided by the average 
energy loss ($dE/dx$) in the tracking devices  and
by an internally reflecting ring-imaging 
Cherenkov detector (DIRC) covering the central region.

%
%
\section{Event Selection} 
\label{sec:presel}

We reconstruct the
$\eta$ mesons in both of the dominant final states $\eta\ra\gamma\gamma$
($\eta_{\gamma\gamma}$) and $\eta\ra\pi^+\pi^-\pi^0$ (\etappp).
For the \etapr , we also reconstruct two final states: \etaptorg\ (\etaprg )
and \etaptoepp\ (\etapepp ), with \etatogg\ (except in the
\fetapfivepiKstz\ mode, where we  
include also \etatoppp ).  The \Kstarz\ is reconstructed as $K^+\pim$ 
(\KstzKppim ), and \Kstarp\ as either $K^+\piz$ 
(\KstpKppiz ) or $\KS\pip$ (\KstpKspip ), with
$\KS\ra\pip\pim$.  The $\rho^+$ is reconstructed as $\pip\piz$.

Monte Carlo (MC) simulations \cite{geant}\ of the signal decay modes and
of continuum and \BB\ backgrounds are used to establish the event selection
criteria.  The selection is designed to achieve high efficiency and
retain sidebands sufficient to characterize the background for
subsequent fitting.  Photons must have energy exceeding a threshold
dependent on the combinatorial background of the specific mode:
$E_\gamma>30$ MeV for the two photons used to reconstruct the \piz\
in $\etatoppp$ candidates, $E_\gamma>$ 100 MeV for \etatogg , and 
$E_\gamma>$ 200 MeV for the photon in \etaprg .
Additionally, in order to reject background from $B\ra K^*\gamma$ 
for the \fetaKst\ and \fetarho\ analyses, we require that 
the cosine of the center of mass decay angle for \etagg\ daughters, relative 
to the flight direction of the $\eta$, have an absolute value of less than 0.86.

We select resonance candidates with the following requirements on the
invariant mass (in \mev) of their final states: $910< m_{\etapr} <1000$
for \etaprg\ and \etapepp , $520< m_{\eta} < 570$ for \etappp,
$755<m_{K^*}<1035$, and $470<m_{\rho^+}<1070$.  For the 
\fetapfivepiKstz\ channel, we tighten the $K^*$ mass range to
$792<m_{K^{*0}}<992$.  Since these quantities are observables input to a
maximum likelihood fit, the criteria are loose.  Additional states are
selected with 2-3 sigma cuts:  $120 < m_{\pi^0} < 150$, $490<
m_{\eta}<600$ for \etagg, tightened to $510< m_{\eta} < 580$
in the \fetapepprhop\ channel to reduce the amount of
continuum background in the sample.  For
$\kzs\ra\pip\pim$ candidates we require $488 < m_{K_S} < 508$.  

In modes with a \KstpKppiz\ or a $\rho^+$, we also require that the
cosine \hel\ of the helicity angle (the vector meson's rest decay angle with
respect to its flight direction) be greater than $-0.5$.  For
\fetapfivepiKstz , we require 
$\hel>-0.9$.  For \kzs\ candidates we require that the lifetime significance
($\tau/\sigma_{\tau}$) be $>3$.

We make several particle identification (PID) requirements to ensure the
identity of the signal pions and kaons.
Tracks in resonance candidates (excluding the \KS ) must have DIRC, 
$dE/dx$, and EMC responses consistent with the expected particle type.  
For the \etappip\ decay, we require that the prompt charged track have
an associated DIRC Cherenkov angle between $-3.5\,\sigma$ and $+3.5\,\sigma$ 
from the expected value for either a pion or a kaon.  The Cherenkov
angle is used in a subsequent fit to distinguish \fetappip\ from 
\fetapKp.

A $B$ meson candidate is characterized kinematically by the energy-substituted mass
$\mes = \sqrt{(\half s + \pvec_0\cdot \pvec_B)^2/E_0^2 - \pvec_B^2}$ and
energy difference $\DE = E_B^*-\half\sqrt{s}$, where the subscripts $0$ and
$B$ refer to the initial \UfourS\ and to the $B$ candidate, respectively,
and the asterisk denotes the \UfourS\ frame. 
The mode-dependent resolutions on these quantities measured for signal
events average about 30 MeV for \DE\ and $3.0\ \mev$ for \mes.
We require $|\DE|\le0.2$ GeV and $5.2\le\mes\le5.29\ \gev$.

\subsection{Tau, QED, and continuum background}

To discriminate against tau-pair and two-photon background we require
that each event contain at least three charged tracks, including at
least one from the recoil $B$ meson in addition to those required to
complete the reconstructed $B$ candidate.

To reject continuum background, we make use of the angle $\theta_T$ between 
the thrust axis of the $B$ candidate and that of
the rest of the tracks and neutral clusters in the event, calculated in
the $\Upsilon(4s)$ frame.  The distribution of $\cos{\theta_T}$ is
sharply peaked near $\pm1$ for combinations drawn from jet-like $q\bar q$
pairs and is nearly uniform for the isotropic $B$ meson decays; we tune the 
\costhr\ cut for each channel to optimize our sensitivity to signal in
the presence of continuum background.  The resulting requirements are
$|\cos{\theta_T}|<0.8$ for \fetaggrhop and \fetapfivepiKstz , 
$|\cos{\theta_T}|<0.75$ for \fetaprgKstz and \fetaprgKstp ,
$|\cos{\theta_T}|<0.65$ for \fetaprgpip and \fetaprgrhop, and
$|\cos{\theta_T}|<0.9$ for all of the other decay sequences.
The average candidate multiplicity is about 1.1 to 1.2 per event.
In events with more than one candidate we accept the one with $\eta$ or
\etapr\ mass closest to the nominal value. 

The remaining continuum background dominates the samples and is modeled
from sideband data for the maximum likelihood fits described in
Section~\ref{sec:mlfit}. 

\subsection{\boldmath \BB\ background}
\label{sec:bbbar}

We use Monte Carlo simulations of \BzBzb\ and \BpBm\ pair production
and decay to look for possible \BB\ backgrounds.  Most \BB\
backgrounds in these analyses come from a handful of charmless decays
with similar final states.  To study these backgrounds, we perform our
preselection cuts on high statistics MC samples of the prominent \BB\
backgrounds.  Using our resulting selection efficiency, along with
measured (or predicted) branching fractions for the charmless decays
in question, we estimate the number of \BB\ background events expected
to enter our samples from each exclusive charmless decay.  From
simulated experiments where we embed the expected number of \BB\
background events from the MC, we determine whether these backgrounds
are large enough to warrant an extra component in the likelihood fit,
as described in Section~\ref{sec:like}.

From these studies we find no evidence for significant \BB\
backgrounds in either the \etatoppp\ decay chains or the
\fetapepppip\ decay.  For the other channels, however, we see evidence
for potential \BB\ backgrounds, normally involving other charmless $B$
decays with a high-momentum $\eta$ or \etapr.  Such backgrounds typically 
are at a level of one or two events and we
include a \BB\ component in the likelihood fits to
discriminate between these backgrounds and the signal.  For those
channels where we include a \BB\ component in the fit, we generate
composite \BB\ background samples, typically containing events from
about five exclusive charmless decay chains.

%
%
\section{Maximum Likelihood Fit}\label{sec:mlfit}

We use an unbinned, multivariate maximum likelihood fit to extract
signal yields for our modes.  A sub-sample of events to fit for each
decay channel is selected as described in Section~\ref{sec:presel}.  The
sample sizes for the decay chains reported here range from 700 to 30000
events. 

\subsection{Likelihood Function} \label{sec:like}

The likelihood function incorporates a number of observables to
distinguish signal from the large number of background events retained
by the sample selection.
We describe the $B$ decay kinematics with two variables: \DE\ and \mes .
We also include the mass of the primary resonance ($m_{\eta}$ or 
$m_{\etapr}$) and a 
Fisher discriminant \xf , which describes energy flow in the event.
The Fisher discriminant combines four
variables: the angles with respect to the beam axis of the $B$ momentum
and $B$ thrust axis (in the \UfourS\ frame), and
the zeroth and second angular moments $L_{0,2}$ of the energy flow
about the $B$ thrust axis.  The moments are defined by
$ L_j = \sum_i p_i\times\left|\cos\theta_i\right|^j,$
where $\theta_i$ is the angle with respect to the $B$ thrust axis of
track or neutral cluster $i$, $p_i$ is its momentum, and the sum
excludes the $B$ candidate.

In addition to these four variables, we include the following
observables relevant to particular decay chains.  For modes with a $K^*$
or $\rho$, we include in the fit the mass and helicity angle of the
vector meson with a two-dimensional probability distribution function (PDF) 
accounting for the different shapes of the helicity-angle distributions for 
true $K^*$ or $\rho$ mesons in the background and for combinatoric background.
To separate \fetappip\ from \fetapKp\ we include in the PDF for the pion
(kaon) signal the observable $S_\pi$ ($S_K$), the Cherenkov
angle residual with respect to the expected angle for pions (kaons)
normalized by the measurement error.

As measured correlations among these observables in the selected
data are small, we take the PDFs for each event $i$ to be a product of the
PDFs for the separate observables.  
We define hypotheses $j$, where $j$ can be signal, continuum
background, or (where appropriate) \BB\ background. The PDF for the
$\eta^{(\prime)}\rho$ and $\eta^{(\prime)}K^*$ analyses is given by

\begin{equation}
{\cal P}^i_{j} =  {\cal P}_j (\mes^i) \cdot {\cal  P}_j (\DE^i) \cdot
 { \cal P}_j(\xf^i) \cdot {\cal P}_j (m^i_{\eta^{(\prime)}}) \cdot {\cal P}_j (m^i_{K^*/\rho},{\cal H}^i).
\end{equation}
For the \fetappip\ analysis, $j$ separately indexes pion and
kaon components:

\begin{equation}
{\cal P}^i_{j} =  {\cal P}_j (\mes^i) \cdot {\cal  P}_j (\DE^i_j) \cdot
 { \cal P}_j(\xf^i) \cdot {\cal P}_j (m^i_{\eta^{(\prime)}}) \cdot {\cal P}_{j} (S^i_j) .
\end{equation}

The likelihood function for each decay mode is

\begin{equation}
{\cal L} = \frac{\exp{(-\sum_j Y_{j})}}{N!}\prod_i^{N}\sum_j Y_{j} {\cal P}^i_{j}\,,
\end{equation}

\noindent where $Y_{j}$ is the yield of events of hypothesis $j$ to be
found by the fitter, and $N$ is the number of events in the sample.  The
first factor takes into account the Poisson fluctuations in the total
number of events.

\subsection{Signal and Background Parameterization}

We determine the PDFs for signal from MC distributions in each
observable.  The PDFs for \BB\ background (where appropriate) arise
from fitting the composite \BB\ MC sample, discussed in
Section~\ref{sec:bbbar}.  For the continuum background we establish
the functional forms and initial parameter values of the PDFs with
data from sidebands in \mes\ or \DE.  We then refine the main background
parameters (excluding resonance mass central values and widths) by
allowing them to float in the final fit. 

The distributions in the resonance mass(es), \mes , and \DE\ for
signal, are parameterized as Gaussian functions, with a second
Gaussian as required for good fits to these samples.  Slowly
varying distributions (combinatoric background under the resonance
mass and \DE\ peaks) are parameterized by linear or quadratic
functions.  We find that two Gaussians also describe the \DE\ shape in
\BB\ background, with one Gaussian typically peaking low (for feed-down) and the
other centered high (for feed-up).  The combinatoric background in \mes\ is
described by a phase-space-motivated empirical (ARGUS) function
\cite{argus}.  For \BB\ background, we use an ARGUS function plus a
Gaussian to parameterize \mes .  We model the \xf\ distribution using
a Gaussian function with different widths above and below the mean; in
background, we include a second Gaussian or a linear contribution to
account for outlying events.  The additional terms ensure that the
background is well modeled in the low-side tail region where signal is found.

The PDFs for the helicity angle variable \hel\ are given by polynomials
(quadratic in signal and 
continuum background, quartic in \BB\ background).  This function is
multiplied by a Fermi-Dirac threshold function
where needed to fit a drop in detector efficiency near $\hel=+1$ or $-1$
where the energy of one of the resonance daughters is low.  As there is a true
resonance component in continuum background, we allow for different
shapes in \hel\ for those events corresponding to real $K^*$ or $\rho$
resonances as opposed to the combinatoric continuum.

The PDFs ${\cal P}(S_{\pi,K})$ are determined with a sample of $D^*$-tagged
$D^0$ decays, and parameterized as double Gaussians.  Consistent with
the calibrations, the kaon PDFs
for $S$ and \DE\ are copies of those for pions evaluated at the
corresponding displaced value of the observable\footnote{To facilitate
simulations and PDF projections we actually work with the transformed
pairs $(S_\pi,\delta S\equiv S_\pi-S_K)$ and $(\DE_\pi,\delta\DE\equiv
\DE_K-\DE_\pi)$}. 

We check the simulation on which we rely for signal PDFs by comparing
with large data control samples.  For \mes\ and \DE\ we use
the decays \Dcontrol, which have similar topology to the modes under study.
For the resonance masses we use inclusive resonance production in data.
We adjust the means and widths of PDF parameterizations based on these
control samples.

%
%

%
%
\section{Fit Results}

By generating (from PDF shapes) and fitting simulated samples of signal
and background, we verify that our fitting procedure is functioning
properly.  We find that the minimum $-\ln{\cal L}$ value in the
on-resonance sample lies well within the $-\ln{\cal L}$ distribution
from these simulated samples.

The efficiency is obtained from the fraction of signal MC events passing
the selection, adjusted for any bias in the likelihood fit.  This bias
is determined from fits to simulated samples, each equal in size to the
data and containing a known number of signal (and \BB, where
appropriate) MC events combined with events generated from the continuum
background PDFs. 
The biases we find depend on the mode, but are a few percent where the
yield is substantial, or at most 2--3 events in the limit of vanishing
yield. 

\begin{table}[htb]
\caption{ Yields, efficiencies, branching fractions, and charge
asymmetries for \B\ decays to states with an $\eta$ meson, measured
with our sample of 89 million \BB\ pairs. The
overall yields and efficiencies ($\epsilon$) of the combined mode are
given as the sum of yields and efficiencies from the various daughter
particle decay channels, and for these we incorporate the systematic
errors.  }
\label{tab:restab_eta}
\begin{center}
\vspace*{-0.3cm}
\hspace*{-0.5cm}
\begin{tabular}{l|ccccccc}
    \dbline
Mode    &Fit \calB$(10^{-6})$&Signif. ($\sigma$)&Signal yield&$\epsilon$ (\%) & \acp	\\
\sgline
\etaKstz	& \retaKstz		&\setaKstz   & 157	   &  9.2	& \AetaKstz\\
~~\fetaggKstz	& $21\pm3$          	& 13.1	& $125\pm16$       & 6.6	& $+0.12\pm0.12$\\
~~\fetapppKstz	& $14\pm4$            	& 6.8	& $32\pm9$         & 2.6	& $-0.39\pm0.25$         \\
\sgline
\etaKstp		& \retaKstp 	&\setaKstp	& 113	   & 4.7 	& \AetaKstp\\
~~\fetaggKstpKspip	& $23\pm6$      & 5.7	& $46\pm12$        & 2.2 	& $+0.04\pm0.24$ \\
~~\fetapppKstpKspip	& $32\pm9$      & 6.2	& $27\pm8$         & 0.9 	& $+0.46^{+0.24}_{-0.28}$\\
~~\fetaggKstpKppiz	& $27\pm8$      & 6.0	& $30\pm9$         & 1.1 	& $-0.06^{+0.26}_{-0.27}$\\
~~\fetapppKstpKppiz	&$20\pm10$      & 4.5	& $10\pm5$         & 0.5 	& $+0.39^{+0.41}_{-0.50}$\\
\sgline					
\etarhop	& \retarhop		&\setarhop	& 65	   & 7.0 	& \Aetarhop\\
~~\fetaggrhop	& $10\pm4$            	& 3.8	& $42\pm17$        & 4.7 	& $+0.46^{+0.34}_{-0.39}$\\
~~\fetappprhop	& $11\pm6$              & 3.0	& $23\pm12$        & 2.3 	& $-0.56^{+0.48}_{-0.24}$\\
\dbline
\end{tabular}
\end{center}
\end{table}

\begin{table}[htb]
\caption{ Yields, efficiencies, and branching fractions for \B\ decays
to states with an $\etapr$ meson, measured with our sample of 89 million
\BB\ pairs.  We quote 90\% CL upper limits if the significance of the
measured yield is less than 4 sigma.  The overall yields and
efficiencies ($\epsilon$) of the combined mode are given as the sum of
yields and efficiencies from the various daughter particle decay
channels, and for these we incorporate the systematic errors.  }
\label{tab:restab_etap}
\begin{center}
\vspace*{-0.3cm}
\hspace*{-0.5cm}
\begin{tabular}{l|cccccc}
    \dbline
Mode    &Fit \calB$(10^{-6})$&UL \calB$(10^{-6})$&Signif. ($\sigma$)&Signal yield&$\epsilon$ (\%) \\
\sgline
\etappip	& \retappip		&$<\uletappip$	&\setappip	& 13		& 9.9 	\\
~~\fetapepppip	& $~4\pm2$            		& 	& 3.9	& $17\pm7$      	& 4.9	\\
~~\fetaprgpip	& $-1\pm3$            		& 	& ---	& $-4\pm10$       	& 5.0	\\
\sgline												  
\etapKstz	& \retapKstz		&$<\uletapKstz$ &\setapKstz	& 22	 	& 6.6	\\
~~\fetapeppggKstz	& $-2\pm2$             	&	& --- 	& $-4\pm4$       & 2.3		\\
~~\fetapfivepiKstz	& $13\pm6$            	& 	& 2.8	& $~11\pm5$       & 1.0		\\
~~\fetaprgKstz		& $5\pm3$            	& 	& 1.9 	& $~~15\pm10$       & 3.4	\\
\sgline												  
\etapKstp	& \retapKstp		&$<\uletapKstp$ &\setapKstp	& 15.4 	       & 3.0	\\
~~\fetapeppKstpKspip	& $-12\pm5$            	&	& ---	& $-8\pm4$             & 0.8	\\
~~\fetaprgKstpKspip	& $~~16\pm9$            & 	& 2.6	& $16\pm9$             & 1.1	\\
~~\fetapeppKstpKppiz	& $~~7\pm7$            	& 	& 1.8	& $~3\pm3$              & 0.5	\\
~~\fetaprgKstpKppiz	&$~~8\pm13$             & 	& 0.7	& $~5\pm7$              & 0.6	\\
\sgline												  
\etaprhop	& ~\retaprhop			&$<\uletaprhop$	&\setaprhop	& 69	 & 4.1	\\
~~\fetapepprhop	& $~11\pm6$      		& 	& 3.1	&$17\pm8$        & 1.8		\\
~~\fetaprgrhop	& $~25\pm12$       		& 	& 2.8	&$52\pm23$       & 2.3		\\
\dbline
\end{tabular}
\end{center}
\end{table}

In Tables~\ref{tab:restab_eta} and~\ref{tab:restab_etap} we show the
results of the fits to the on-resonance data.  Shown for each decay mode
are the measured branching fraction and the 90\% confidence level upper
limit (where appropriate).  We report the statistical significance for
the individual decay chains and display the significance including
systematics for the combined result in each channel.  The statistical
significance is taken as the square root of the difference between the
value of $-2\ln{\cal L}$ for zero signal and the value at its
minimum. The tables also include the fit signal yield, where the
statistical error on the number of events is taken as the change in the
central value when the quantity $-2\ln{\cal L}$ changes by one unit.
The 90\% C.L. upper limit is taken as the solution $B$ to the condition
$\int_0^B{\cal L}(b)db/\int_0^\infty{\cal L}(b)db=0.9$. The overall
efficiency is given by the product of the Monte Carlo efficiency,
efficiency corrections, and branching fraction products for the daughter
particle decay sequences.  The number of produced $B$ mesons is computed
with the assumption of
equal production rates of charged and neutral pairs.  The final column
in Table~\ref{tab:restab_eta} gives the charge asymmetry
(\acp ).

\begin{figure}[htbp]
\vspace{0.5cm}
 \includegraphics[angle=0,width=\linewidth]{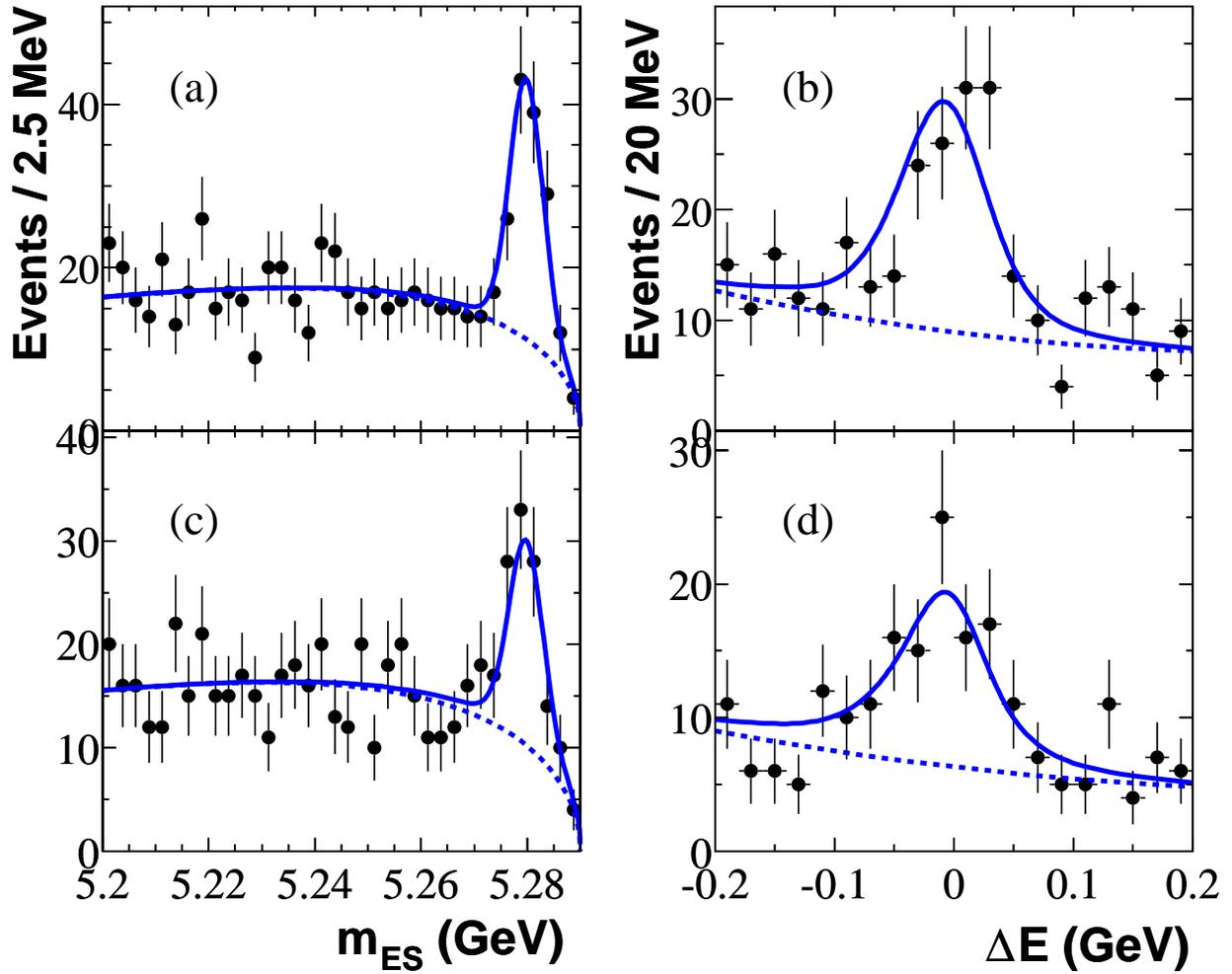}
 \caption{\label{fig:proj_etakstar} Projections of the $B$ candidate
 \mes\ and \DE\ for \fetaKstz\ (a, b) and \fetaKstp\ (c, d). Points with
 errors represent data, 
solid curves the full fit functions, and dashed curves
 the background functions.  
These plots are made with a cut on
 the signal likelihood and thus do not show all events in the data
 samples.  }
\end{figure}

\begin{figure}[htbp]
\scalebox{1.0}[.9]{ \includegraphics[angle=0,width=\linewidth]{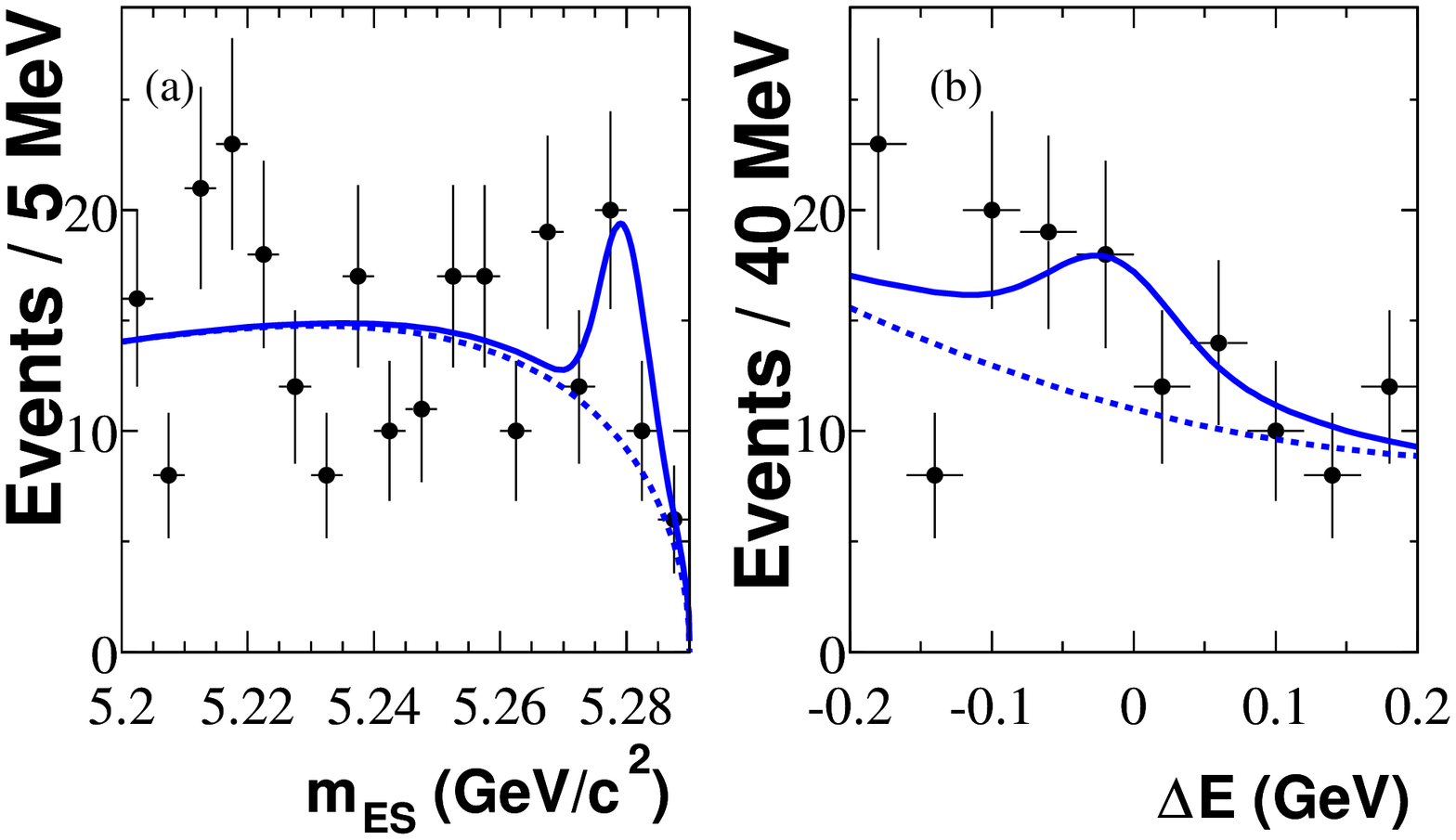}}
\vspace{-2cm}
 \caption{\label{fig:proj_rho} Projections of the $B$ candidate \mes\ (a)
 and \DE\ (b) for \etarhop. Points with
 errors represent data, 
solid curves the full fit functions, and dashed curves
 the background functions.  
These plots are made with a cut on
 the signal likelihood and thus do not show all events in the data
 samples.  }
\end{figure}

In Figures~\ref{fig:proj_etakstar} and~\ref{fig:proj_rho}, we show
projections of \mes\ and \DE\ for the \fetaKst\ and 
$\eta\rho^+$ modes, respectively.  
We make these plots by selecting events with signal likelihood (computed
without the variable shown in the figure) exceeding a mode-dependent
threshold that optimizes the expected sensitivity.  The selection
retains a fraction of the signal yield averaging about 70\% across the
decay sequences.

%
%
\section{Systematic Uncertainties}
\label{sec:syst}

Floating background parameters in the fit allows us to
incorporate into the overall statistical error most of the systematic
errors on yields that arise from uncertainties in the values of the PDF
parameters.  We determine the sensitivity to parameters of the signal
PDF components by varying these within their uncertainties in all
channels except for \fetappip . In the \fetappip\ decay, we float the
primary signal parameters and determine the effect on signal yield; this
is possible because of the simultaneous fit with \fetapKp , which has a
large branching fraction.  This is the only systematic error on the fit
yield; the other systematics apply to either the efficiency or the
number of \BB\ events.

The uncertainty in our knowledge of the efficiency is found to be
0.8$N_t$\%, 2.5$N_\gamma$\%, and 3\%\ for a \KS\ decay, where $N_t$
and $N_\gamma$ are the number of signal tracks and photons,
respectively.  We estimate the uncertainty in the number of produced
\BB\ pairs to be 1.1\%.  The estimated error on systematic bias from the
fitter itself (1--12\%) comes from fits of simulated samples with
varying signal and background populations.  Published world averages
\cite{PDG2002}\ provide the $B$ daughter branching fraction
uncertainties.
We account for systematic effects in \costhr\ ($1-3.5\%$, depending
upon how tight a cut is made), in the PID requirements (2\%), and from
MC statistics (1.2\%).
For decay channels where a
\BB\ component is included in the fit, we assign an additional systematic 
determined by varying the \BB\ yield within its errors.  We quote the 
\BB\ background systematic to be one half the measured change in the signal 
yield when the number of fit \BB\ events is changed by one standard deviation.
For the \fetaprhop\ decay, where the fit \BB\ background yield is about 
five times larger than other decays, we also vary the \BB\ background
model to account for different amounts of the expected individual backgrounds.

A study of the charge asymmetry as a function of momentum for all tracks in 
hadronic events bounds the tracking efficiency component of charge-asymmetry 
bias to be less than 1\%.  $D^*$-tagged $D\ra K\pi$ and $B$ samples provide
additional crosschecks that the bias is small.
We assign a systematic uncertainty for \acp\ of 2\% based on the tracking 
study and a small PID contribution determined from the $D^*$ studies.

We keep track of which systematic error contributions are (un)correlated
between the several measurements with different secondary decay modes of
the same primary decay for use in obtaining their average as the final
branching fraction result.

\section{\boldmath Combined Results}
\label{sec:combine}

\begin{figure}[htbp]
\scalebox{1.0}{ \includegraphics[width=\linewidth]{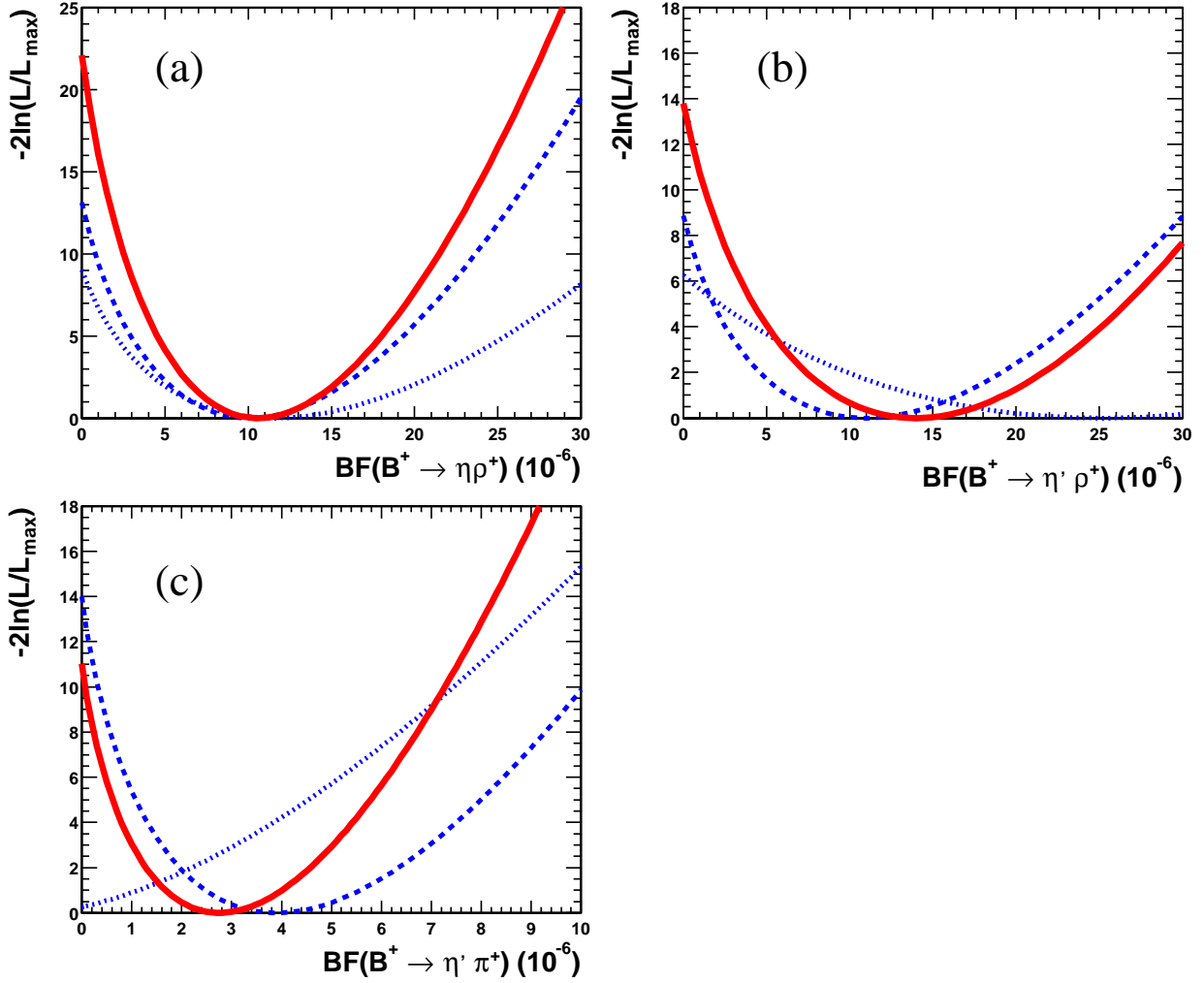}}
\caption{\label{fig:nll} Distributions of $-2\ln{\cal L}$ convolved
with uncorrelated systematic errors for branching fraction
measurements.  Solid curves represent the result of combining
channels.  The frames present (a) \fetarhop, with \fetaggrhop\
(dashed) and \fetappprhop (dotted); (b) \fetaprhop, with
\fetapepprhop\ (dashed) and \fetaprgrhop\ (dotted); and (c) \fetappip,
with \fetapepppip\ (dashed) and \fetaprgpip\ (dotted).  } 
\end{figure}

To obtain the final results we combine the branching fraction and charge
asymmetry measurements from the individual daughter decay chains.  The
joint likelihood is given by the product, or equivalently $-2\ln\calL$
is given by the sum, of contributions from the submodes.  The
statistical contribution comes directly from the likelihood fit, which
reflects the non-Gaussian uncertainty associated with small statistics.
Before combining we convolve each statistical $-2\ln\calL$ with a Gaussian
function representing the part of the systematic error that is
uncorrelated among the submodes.  We show the resulting distributions in Fig.\
\ref{fig:nll}\ for our measurements of previously unseen decays that
have significance greater than 3 sigma.  The corresponding distributions
without systematics give the combined statistical errors, and these in
conjunction with the solid curves in Fig.\ \ref{fig:nll}\ and the
correlated systematics give the total systematic errors.

The resulting branching fractions and charge asymmetries are included
in Tables~\ref{tab:restab_eta} and~\ref{tab:restab_etap}, where the
significance given includes systematics.

%
%
\section{Conclusion}
\label{sec:conclusion}

We report preliminary measurements of branching fractions and \acp\ for
$B$ meson decays to $\eta$ or \etapr\ with a $K^*$, $\rho^+$, or $\pip\!$.
We find signals with statistical significance exceeding four standard
deviations in all $\eta$ channels. The decay \etarhop\ has not been seen
previously.  We also have evidence for \etaprhop\ (with significance
$\setaprhop\sigma$) and \etappip\ (with significance $\setappip\sigma$).
The observed values in the $\eta$ channels are
\begin{eqnarray*}
\BetaKstz &=& \RetaKstz\,, \\
\BetaKstp &=& \RetaKstp\,, \\
\Betarhop &=& \Retarhop\,.
\end{eqnarray*}
For the \etapr\ channels we find
\begin{eqnarray*}
\Betappip &=& \Retappip\ (<\ULetappip), \\
\BetapKstz &<& \ULetapKstz\,, \\
\BetapKstp &<& \ULetapKstp\,, \\
\Betaprhop &=& \Retaprhop\ (<\ULetaprhop),
\end{eqnarray*}
where the upper limits are taken at 90\% CL.   These results supersede
the previous \babar\ 
measurements~\cite{BABARetapOmega,BABAReta}.  They represent substantial
improvements over all previous measurements, as can be seen from Tables
\ref{tab:OldResultsa}\ and \ref{tab:OldResultsb}.  The branching fraction
limit for \etappip\ is nearly three times more restrictive than previous
measurements. 
The measurement for \etapKst\ is not yet precise enough to determine 
whether a flavor singlet component is present for this decay, though we
do restrict the size of this contribution.

For the modes with significant signals, we measure the charge asymmetries
\begin{eqnarray*}
\acp(\fetaKstz) &=& \AetaKstz\,, \\
\acp(\fetaKstp) &=& \AetaKstp\,, \\ 
\acp(\fetarhop) &=& \Aetarhop\,. 
\end{eqnarray*}
These charge asymmetry
results are in agreement with the theoretical expectations discussed
in Section \ref{sec:intro}\ and rule out substantial portions of the physical region.

\section{Acknowledgments}
\label{sec:Acknowledgments}


We are grateful for the 
extraordinary contributions of our \pep2\ colleagues in
achieving the excellent luminosity and machine conditions
that have made this work possible.
The success of this project also relies critically on the 
expertise and dedication of the computing organizations that 
support \babar.
The collaborating institutions wish to thank 
SLAC for its support and the kind hospitality extended to them. 
This work is supported by the
US Department of Energy
and National Science Foundation, the
Natural Sciences and Engineering Research Council (Canada),
Institute of High Energy Physics (China), the
Commissariat \`a l'Energie Atomique and
Institut National de Physique Nucl\'eaire et de Physique des Particules
(France), the
Bundesministerium f\"ur Bildung und Forschung and
Deutsche Forschungsgemeinschaft
(Germany), the
Istituto Nazionale di Fisica Nucleare (Italy),
the Foundation for Fundamental Research on Matter (The Netherlands),
the Research Council of Norway, the
Ministry of Science and Technology of the Russian Federation, and the
Particle Physics and Astronomy Research Council (United Kingdom). 
Individuals have received support from 
the A. P. Sloan Foundation, 
the Research Corporation,
and the Alexander von Humboldt Foundation.

\end{document}